\documentclass[review]{elsarticle}

\usepackage{bm,amssymb,amsmath,mathrsfs}
\usepackage{amsthm}
\usepackage{lineno}
\usepackage[usenames]{color}

\usepackage{dcolumn}

\theoremstyle{plain}

\newtheorem*{theorem*}{Theorem}

\begin{document}

\journal{(internal report CC26-5)}

\begin{frontmatter}

\title{Comment on ``Regarding the Rotational Unruh Effect''}

\author[cc]{S.~R.~Mane}
\ead{srmane001@gmail.com}
\address[cc]{Convergent Computing Inc., P.~O.~Box 561, Shoreham, NY 11786, USA}

\begin{abstract}
We comment on various statements in a recent document 
(A.~Deur, S.~J.~Brodsky, C.~D.~Roberts and B.~Terzi{\'c},
Regarding the Rotational Unruh Effect,
\textit{arXiv:2607.25004v1 [hep.ph]}, (2026)).
The examples cited treat electrons (or positrons) circulating and emitting photons in high-energy storage rings.
The topic is also of interest in astrophysics, for electrons orbiting in magnetic fields around neutron stars.
\end{abstract}

\begin{keyword}
  high-energy storage rings
  \sep synchrotron radiation
  \sep electromagnetic radiation astronomy
  \sep Unruh effect
  \sep QED
\end{keyword}

\end{frontmatter}

\setcounter{equation}{0}
\section{Introduction}\label{sec:intro}
Deur, Brodsky, Roberts and Terzi{\'c} \cite{RUE} recently published a note on the ``Rotational Unruh Effect'' 
(denoted `RUE' in \cite{RUE}).
Unruh \cite{Unruh} discovered that an observer in a uniformly (linear) accelerating reference frame
observes that the vacuum electromagnetic fluctuations have a thermal (blackbody) spectrum, with a temperature proportional to the proper acceleration.
The rotational (or circular) Unruh effect pertains to the case where the observer moves in a circle.

We comment on various statements in \cite{RUE}.
In particular, some of their descriptions of prior work by other authors \cite{BL1987,ST,JacksonRMP} are not completely accurate.
We focus on the case where the observer is an electron, and it emits photons,
i.e., the observer is a detector which generates an experimentally observable signal.
A significant example is the work of Bell and Leinaas \cite{BL1987,BL1983}.
We also cite the Sokolov-Ternov effect \cite{ST}, where electrons or positrons circulating in high-energy storage rings
become spontaneously polarized by the emission of spin-flip synchrotron radiation.
We explain why the asymptotic polarization level is lower than $100\%$.
We also cite the work of Hacyan and Sarmiento \cite{HS}, who calculated the vacuum stress-energy tensor of the electromagnetic field in a uniformly rotating frame
and found there is a nonzero energy flux in the direction of motion of the observer in such a frame.
Our notation is standard: the speed of light is $c$ and we treat a particle of mass $m$, charge $e$, velocity $\bm{v}=\bm{\beta}c$
and Lorentz factor $\gamma=1/\sqrt{1-\beta^2}$. We also define $a=(g-2)/2$ for the spin.
We also retain $\hbar$ explicitly, to clarify ``quantum'' as opposed to ``classical'' effects.

\setcounter{equation}{0}
\section{Sokolov-Ternov effect}\label{sec:ST}
In 1964, Sokolov and Ternov \cite{ST} showed that the spins of electrons and positrons circulating in a high-energy storage ring
become spontaneously polarized by the emission of spin-flip synchrotron radiation.
This phenomenon is now called the ``Sokolov-Ternov effect'' in their honor.
They derived that the asymptotic polarization level, say $P_{\rm ST}$, is
\begin{equation}
\label{eq:P_ST}
P_{\rm ST} = \frac{8}{5\sqrt3} \simeq 0.924 \,.
\end{equation}
Quoting from (\cite{RUE}, Section III A):
\begin{quote}
It is well-established and experimentally verified that there
is a theoretical limit for the natural polarization of electrons in
storage rings. This is the Sokolov-Ternov effect [14] \emph{(Ref.~\cite{ST} in this note)}. It arises
when particles, in practice electrons or positrons, emit synchrotron
radiation. The effect has a theoretical maximum polarization
of $P_{\rm max} = 8/5\sqrt3 \simeq 0.924$.
This limit was first derived
semi-classically [14] \emph{(Ref.~\cite{ST} in this note)} and confirmed by subsequent studies
[15, 16] \emph{(Refs.~\cite{DK73} and \cite{JacksonRMP} respectively in this note)}.
All these studies were conducted before the Unruh
effect was suggested [3] \emph{(Ref.~\cite{Unruh} in this note)}.
\end{quote}
The above statements must be interpreted with care.
\begin{itemize}
\item
The term ``theoretical maximum polarization'' probably means the \emph{asymptotic} polarization level.
For example, in the current design of the Electron-Ion Collider at Brookhaven National Laboratory,
electrons from a polarized electron source will be injected into the electron ring,
and the spin-flip synchrotron radiation will \emph{decrease} the polarization of the stored electron beam to an asymptotic value.
See, e.g.,~\cite{SH_BAGELS2} for an analysis of the subject.
We shall employ the term ``asymptotic polarization level'' (or equivalently ``equilibrium polarization level'') in this note.
\item
Next, Sokolov and Ternov \cite{ST} calculated the synchrotron radiation by employing the Dirac equation with a static uniform vertical magnetic field,
not a semiclassical formalism.
They calculated transitions between eigenstates of the Dirac Hamiltonian, which are now called ``Sokolov-Ternov states'' in their honor.
They treated only the case $g=2$ (i.e., no anomalous magnetic moment).
They derived the asymptotic polarization level stated in eq.~\eqref{eq:P_ST} above.
\item
Derbenev and Kondratenko \cite{DK73} employed a semiclassical formalism.
They treated a model of motion in inhomogeneous accelerator fields, e.g., quadrupole magnets.
The electrons executed bounded orbital oscillations around a reference orbit.
Their formalism also treated arbitrary values of $g$.
\item
In 1976, Jackson \cite{JacksonRMP} published a by-now classic review of spin-flip synchrotron radiation in high-energy storage rings,
also employing a semiclassical formalism and arbitrary values of $g$.
Jackson's model treated only electron motion along the reference orbit.
\end{itemize}

\setcounter{equation}{0}
\section{Asymptotic polarization level lower than $100\%$}\label{sec:Plt1}
In his review, Jackson \cite{JacksonRMP} noted that the electrons in a storage ring,
in particular the idealized model of horizontal circular motion in a static uniform vertical magnetic field,
are essentially at zero temperature.
Hence why is the asymptotic polarization level lower than $100\%$?
Quoting from (\cite{JacksonRMP}, Section VI):
\begin{quote}
  The reader may, with justification, feel that the author has wandered endlessly in a labyrinth of Airy functions without coming to grips with the minotaur,
  the mysterious and peculiar $8/5\sqrt{3}$\,! Why \emph{is} the polarization for electrons so large, and yet not complete? I have no compelling answer.
\end{quote}
I sent Jackson the following answer.
We first note that electrons orbit in the magnetic field of a neutron star and emit photons
and the topic is of interest in astrophysics (where the phenomenon is termed ``cyclotron radiation'').
See early work by Latal \cite{Latal}, also the later papers \cite{BaringAstro} and \cite{SemionovaAstro}
(this does not pretend to be a complete bibliography of the astrophysics literature on cyclotron radiation).
For an electron orbiting a neutron star, the electron's orbit decays as photons are emitted, and the electron eventually drops to the ground state,
which is non-degenerate (because of the magnetic field).
Hence the asymptotic spin polarization level is indeed $100\%$.
However, for an electron circulating in a storage ring,
the ring contains radio-frequency cavities which replenish the energy loss of the electron due to the emission of photons.
In the models treated in the HEP literature, an electron circulating in a storage ring can radiate indefinitely, and its orbit does not decay.
An electron emitting a photon in a storage ring can therefore always find a lower energy state of the opposite spin orientation.
Hence the asymptotic electron spin polarization level in a storage ring is lower than $100\%$.
Jackson \cite{JDJprivcomm} kindly accepted my explanation as a satisfactory answer to his challenge.

It is implicit in most of the HEP literature, e.g., the review by Jackson \cite{JacksonRMP},
the analysis by Bell and Leinaas \cite{BL1987,BL1983} (see below), and also the note by
Deur, Brodsky, Roberts and Terzi{\'c} \cite{RUE}, that an electron's orbit does not decay despite the electron radiating photons indefinitely.
However, the astrophysics literature is cognizant that photon emissions cause an electron's orbit to decay and eventually reach the ground state.

\setcounter{equation}{0}
\section{Asymptotic polarization level for arbitrary $g$}\label{sec:P_a}
\begin{figure}[!h]
\centering
\includegraphics[width=0.9\textwidth]{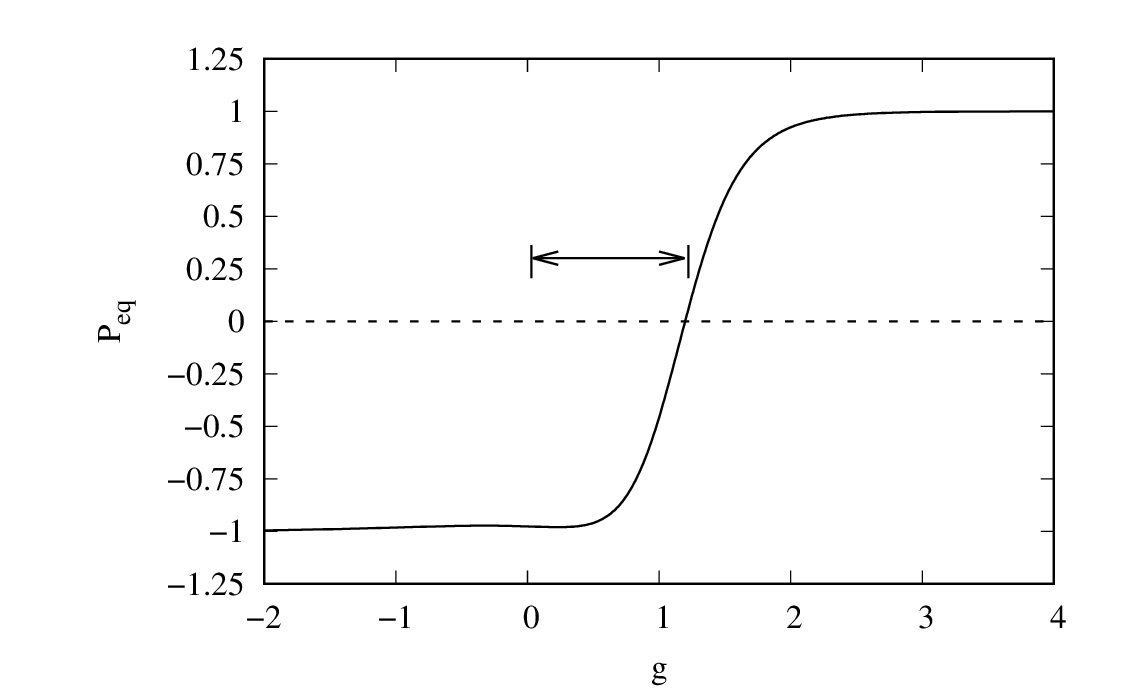}
\caption[\;Graph of asymptotic polarization vs.~$g$]{
\label{fig:graph_pol}
Graph of the asymptotic polarization $P_{\rm eq}$ as a function of $g$, for a model of horizontal circular motion in a static uniform vertical magnetic field.
For $0 < g \lesssim 1.2$, the polarization is negative, i.e.~the na{\i}vely higher-energy spin state is preferentially populated.
The range is indicated by the horizontal arrow.}
\end{figure}

Both Derbenev and Kondratenko \cite{DK73} and Jackson \cite{JacksonRMP} employed semiclassical formalisms 
to treat a model of electrons executing horizontal circular motion in a static uniform vertical magnetic field, but with an arbitrary value of $g$.
They obtained the same answer. For $g=2$, their results both equal the value in eq.~\eqref{eq:P_ST}.
Recall $a=(g-2)/2$.
The expressions below follow Jackson (\cite{JacksonRMP}, eq.~(61)):
\begin{equation}
\begin{split}
  F_1(a) &= 1 +\frac{41}{45}a -\frac{23}{18}a^2 -\frac{8}{15}a^3 +\frac{14}{15}a^4
  -\frac{8}{5\sqrt3}\frac{a}{|a|}\Bigl(1 +\frac{11}{12}a -\frac{17}{12}a^2 -\frac{13}{24}a^3 +a^4\Bigr) \,,
\\
F_2(a) &= \frac{8}{5\sqrt3}\Bigl(1 +\frac{14}{3}a +8a^2 +\frac{23}{3}a^3 +\frac{10}{3}a^4 +\frac{2}{3}a^5\Bigr) \,.
\end{split}
\end{equation}
The asymptotic (or equilibrium) polarization level is (\cite{JacksonRMP}, eq.~(63))
\begin{equation}
\label{eq:P_a}  
  P_{\rm eq}(a) = \frac{F_2(a)}{\displaystyle F_1(a)e^{-\sqrt{12}|a|} +\frac{a}{|a|}F_2(a)} \,.
\end{equation}
A graph of $P_{\rm eq}(a)$ is plotted as a function of $g$ in Fig.~\ref{fig:graph_pol}.
Observe that $P_{\rm eq}\simeq-0.98$ for $g=0$, i.e., the asymptotic polarization level is nonzero even though the magnetic moment is zero.
Also, $P_{\rm eq}<0$ for $0\le g \lesssim 1.2$, i.e., the na{\"\i}vely higher energy spin state is preferentially populated.
These facts were noted by Jackson \cite{JacksonRMP}.

Deur, Brodsky, Roberts and Terzi{\'c} (\cite{RUE}, Section III B) presented a derivation of the Sokolov-Ternov effect.
They employed a Dirac Hamiltonian with a Pauli anomalous magnetic moment term, plus additional terms due to the rotating reference frame (\cite{RUE}, eq.~(5)).
Hence their formalism includes an arbitrary $g$-factor (see \cite{RUE}, eqs.~(6), (17), (B), (C), (19), (25), (31) and (45)).
However, they did not publish a formula for the asymptotic polarization level as a function of $g$, which can be compared to eq.~\eqref{eq:P_a},
i.e., there is no direct comparison of the equivalence of their formalism to that of Derbenev and Kondratenko \cite{DK73} and Jackson \cite{JacksonRMP}.

\setcounter{equation}{0}
\section{Bell-Leinaas model}\label{sec:BL}
Bell and Leinaas \cite{BL1987,BL1983} were interested in finding an observable signal to detect the Unruh effect.
The acceleration attainable in linear motion is too small to be useful, but much larger accelerations can be attained in circular motion.
Hence Bell and Leinaas studied the motion of electrons circulating in a storage ring.
For an experimentally detectable signal, they studied the coupling of the electron spin to the photon field, i.e., spin-flip photon emissions,
and calculated the resulting asymptotic electron spin polarization.
Deur, Brodsky, Roberts and Terzi{\'c} \cite{RUE} cited the work of Bell and Leinaas \cite{BL1987,BL1983}.
Quoting from (\cite{RUE}, Section III B):
\begin{quote}
The result of Ref.~[13] \emph{(Ref.~\cite{BL1987} in this note)}, derived in the comoving frame,
agrees with that of Ref.~[16] \emph{(Ref.~\cite{JacksonRMP} in this note)}, derived in the laboratory frame with cylindrical symmetry.
In the latter, the classical Thomas precession causes $P_{\rm max} \simeq 0.924$.
\end{quote}
Note the following.
\begin{itemize}
\item
The asymptotic polarization formula derived by Bell and Leinaas \cite{BL1987} is different from that derived by Jackson \cite{JacksonRMP}.
The Bell-Leinaas formula exhibits a spin resonance (see eq.~\eqref{eq:P_BL} below)
whereas that by Jackson does not (see eq.~\eqref{eq:P_a} above).
The concept of a spin resonance will be explained below.

\item
The reason for the difference in formulas is that the two sets of authors treated different models.
Bell and Leinaas \cite{BL1987} treated a weak-focusing storage ring:
the electrons executed (bounded) vertical oscillations, induced by vertical momentum recoils due to photon emissions.
Jackson \cite{JacksonRMP} treated only electron motion on the reference orbit;
his model had no accelerator focusing fields and no orbital oscillations around the reference orbit.

\item
As noted in eq.~\eqref{eq:P_a} above, Jackson \cite{JacksonRMP} (also Derbenev and Kondratenko \cite{DK73})
calculated the asymptotic polarization level for all values of $g$, not just $g=2$
(for a model of electrons executing horizontal circular motion in a static uniform vertical magnetic field).
The asymptotic polarization level is not always $8/(5\sqrt3) \simeq 0.924$ in Jackson's analysis.
\end{itemize}

\begin{figure}[!h]
\centering
\includegraphics[width=0.9\textwidth]{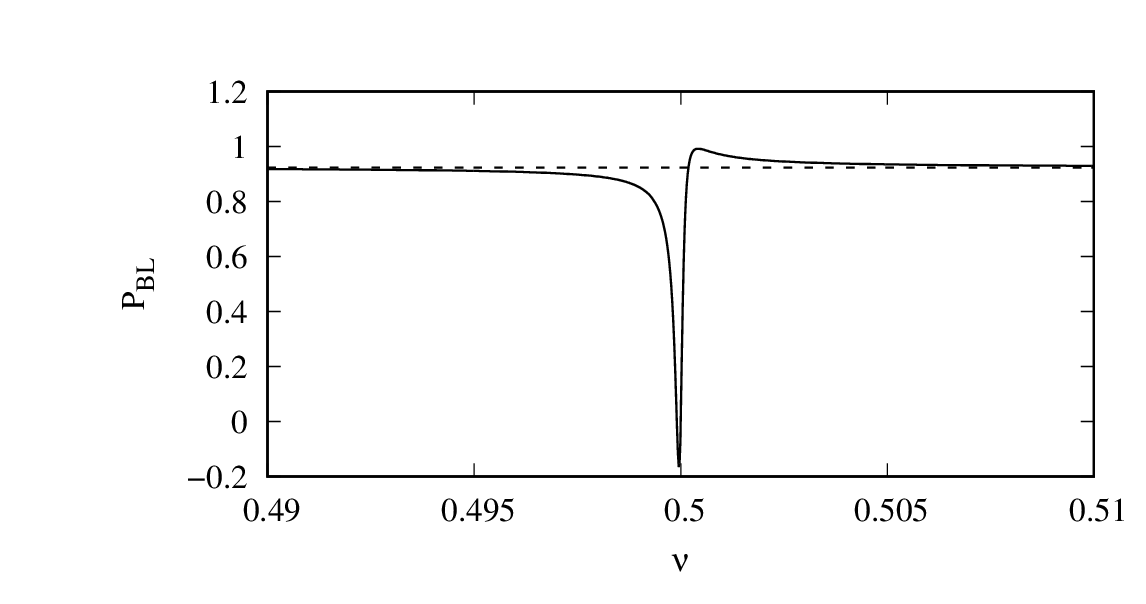}
\caption[\;Graph of asymptotic polarization vs.~$\nu$]{
\label{fig:graph_pol_BL}
Graph of the asymptotic polarization $P_{\rm BL}$ as a function of $\nu$,
for the model treated by Bell and Leinaas \cite{BL1987}.
Here $P_{\rm BL}$ is the asymptotic polarization level and $\nu=\gamma a$, where $a=(g-2)/2$.
The dashed line indicates the Sokolov-Termov polarization value $8/(5\sqrt3) \simeq 0.924$.}
\end{figure}

Barber and Mane \cite{BM} employed inertial-frame QED and derived the asymptotic polarization formula by Bell and Leinaas \cite{BL1987}.
Recall that it was stated above that
Derbenev and Kondratenko \cite{DK73} employed a semiclassical formalism and treated a model of motion in inhomogeneous accelerator fields.
The electrons executed bounded orbital oscillations around a reference orbit.
The electron motion in high-energy storage rings is ultrarelativistic,
hence Derbenev and Kondratenko \cite{DK73} treated only the longitudinal momentum recoil when an electron emits a photon.
Barber and Mane \cite{BM} extended Derbenev and Kondratenko's formalism to also include vertical momentum recoils.

The exposition below follows Barber and Mane \cite{BM}.
The radius of the reference orbit is $R$.
The ``tune'' of the vertical orbital oscillations is denoted by $Q$.
The tune is a dimensionless frequency, defined as the orbital oscillation frequency divided by the circulation frequency around the ring circumference.
The angular circulation frequency is $\omega_0 = \beta c/R$.
Also $\nu = \gamma a$, where recall $a=(g-2)/2$.
The term ``weak-focusing'' means the focusing fields are cylindrically symmetric around the vertical axis.
Hence the overall accelerator model is cylindrically symmetric around the vertical axis.
Then (\cite{BM}, eq.~(16))
\begin{equation}  
\label{eq:f_BL}
  f = \frac{(g-2)Q^2}{Q^2-\nu^2} \,.
\end{equation}  
The asymptotic polarization formula by Bell and Leinaas, say $P_{\rm BL}$, is (\cite{BL1987}, eq.~(B1), also \cite{BM}, eq.~(17))
\begin{equation}
\label{eq:P_BL}
  P_{\rm BL} = \frac{8}{5\sqrt3}\frac{\displaystyle 1 -\frac{f}{6}}{\displaystyle 1 -\frac{f}{18} +\frac{13}{360}f^2} \,.
\end{equation}  
A graph of the asymptotic polarization level $P_{\rm BL}$ vs.~$\nu$ is plotted in Fig.~\ref{fig:graph_pol_BL}, for $Q=0.5$.
(The dashed line indicates the Sokolov-Ternov value of $8/(5\sqrt3)\simeq 0.924$.)
The asymptotic polarization level exhibits a spin resonance: observe that the value of $f$ in eq.~\eqref{eq:f_BL} diverges at $\nu=Q$.
A spin resonance is caused by spin-orbit coupling of the spin precession
to a linear combination of orbital oscillations around the reference electron orbit.
The asymptotic polarization level equals zero at the center of a spin resonance.
Far from the spin resonance, the polarization level approaches the Sokolov-Ternov value of $8/(5\sqrt3)\simeq 0.924$.
Close to the spin resonance, the asymptotic polarization level drops to a minimum value of approximately $-0.169$ and reaches a maximum value of approximately $0.992$.
From \cite{BM}, the extremum values of $P_{\rm BL}$ do not depend on the value of $Q$, but the width of the spin resonance increases as the value of $Q$ increases.

Hence, even for a model accelerator with cylindrical symmetry,
the contribution of the orbital oscillations (induced by recoils from photon emissions)
can cause the asymptotic polarization level to depend on parameters such as the beam energy (recall $\nu \propto \gamma$)
and the focusing properties of the storage ring (the value of $Q$ in this case),
and phenomena such as spin resonances can occur.
The asymptotic polarization level is not simply a constant.

For the record,
Barber and Mane also published a formula for the asymptotic polarization level treating vertical momentum recoils in a planar \emph{strong-focusing} storage ring (\cite{BM}, eq.~(41)).
Such a model does not have cylindrical symmetry around the vertical axis.

\setcounter{equation}{0}
\section{Hacyan-Sarmiento model}\label{sec:HS}
Hacyan and Sarmiento \cite{HS} calculated the vacuum stress-energy tensor of the electromagnetic field in a rotating frame.
Quoting them (\cite{HS}, Section I):
\begin{quote}
The aim of the present paper is to calculate explicitly
the vacuum stress-energy tensor of the electromagnetic
field as detected in a uniformly rotating frame.
\end{quote}
Hacyan and Sarmiento derived that there is a nonzero energy flux in the direction of motion of the observer in such a frame.
Quoting them (\cite{HS}, Section V):
\begin{quote}
If this flux is real, it should imply some friction-like effect on a rotating particle.
\end{quote}
Hacyan and Sarmiento employed natural units, where $\hbar=c=1$, and obtained the following expression for the energy flux, directed along the tangent to the orbit,
(\cite{HS}, eq.~(4.2b)):
\begin{equation}
\label{eq:HS_energy_flux}
p = \frac{\gamma^4\Omega^4v}{720\pi^2}(50 -47\gamma^{-2}) \,.
\end{equation}
Mane \cite{ManeHS} pointed out that the friction-like effect on a rotating particle is indeed real and has been observed as classical synchrotron radiation \cite{Schwinger1949}.
From (\cite{HS}, eq.~(4.13)), the Poynting four-vector is $pn^\mu$, where $n^\mu =(v, -\sin(\Omega\tau), \cos(\Omega\tau), 0)$,
$\Omega$ is the angular revolution frequency, $\tau$ is the proper time and the Minkowski metric signature is $(+,-,-,-)$.
It is helpful to display $\hbar$ and $c$ explicitly in this note.
The time component of the energy flux is (\cite{ManeHS}, eq.~(2), neglecting the term $\gamma^{-2}$ in eq.~\eqref{eq:HS_energy_flux}, because we assume $\gamma\gg1$)
\begin{equation}
\label{eq:Mane_comment_eq2}
pn^0 \propto \frac{\hbar}{c^4}\gamma^4\Omega^4v^2 \,.
\end{equation}
Observe that the energy flux is proportional to $\hbar$.
To obtain the radiated power by a particle such as an electron,
the relevant coupling factor is the electromagnetic fine-structure constant $\alpha \propto e^2/(\hbar c)$.
Then radiated power per unit area per unit time is
\begin{equation}
\label{eq:Mane_comment_eq3}
\alpha pn^0 \propto \frac{e^2}{c^5}\gamma^4\Omega^4v^2 \,.
\end{equation}
Quoting from \cite{ManeHS}:
\begin{quote}
Note that this \emph{($\alpha pn^0$ above)} is \textit{independent of $\hbar$}.
This explains why the Hacyan-Sarmiento energy flux, which is a quantum effect
proportional to $\hbar$, can be related to classical synchrotron radiation.
However, one should recognize that this comparison is valid only to the leading order in perturbation theory.
As shown by Schwinger${}^8$ \emph{(Ref.~\cite{Schwinger1954} in this note)}, the synchrotron radiation
power spectrum can be expanded in a power series in $\hbar$, where the leading term is classical, i.e., independent of $\hbar$.
In the Hacyan-Sarmiento case, they treat the observer
classically, by specifying a classical world line $x(\tau)$, although they quantize the electromagnetic field.
\end{quote}
Mane \cite{ManeHS} also processed the other components of the Poynting four-vector and showed that the radiated power, which is Lorentz invariant,
matched the analysis using Hacyan and Sarmiento's formalism \cite{HS}.

Deur, Brodsky, Roberts and Terzi{\'c} \cite{RUE} did not cite Hacyan and Sarmiento's work \cite{HS}.
This section is included to indicate to the reader that there are additional calculations of
vacuum effects in circularly accelerating reference frames, not necessarily involving the electron spin.
The results yield observable effects and match an analysis using inertial-frame QED.

\setcounter{equation}{0}
\section{Conclusion}\label{sec:conc}
To date, theoretical calculations to detect signatures of the Unruh effect in circularly accelerating reference frames,
e.g., the work of Bell and Leinaas \cite{BL1987,BL1983} and by Hacyan and Sarmiento \cite{HS}, have been reproduced by other authors using inertial-frame QED.
See, e.g., the analyses in \cite{BM} and \cite{ManeHS}.
It has also been explained why the asymptotic spin polarization level of electrons circulating in a storage ring is lower than $100\%$.
The connection to cyclotron radiation in astrophysics was also pointed out.



\end{document}